%% file: 00_main.tex
\documentclass{article}

\input{0_preamble}

\title{MOTO: Topology Optimization for Large Deformations via an Implicit Material Point Method}

\author{ \and
Rahul Kumar Padhy \\
Department of Mechanical Engineering\\
University of Wisconsin-Madison\\
Madison, WI, USA \\
\texttt{rkpadhy@wisc.edu}
\and 
 Aaditya Chandrasekhar \\
 Department of Mechanical Engineering\\
 Northwestern University\\
 Evanston, IL, USA \\
\texttt{aadityacs@northwestern.edu} \\
\and 
Krishnan Suresh\\
Department of Mechanical Engineering\\
University of Wisconsin-Madison\\
Madison, WI, USA \\
\texttt{ksuresh@wisc.edu}
}

\begin{document}
\maketitle

\begin{abstract}

The Finite element method (FEM) has long served as the computational backbone for topology optimization (TO). However, for designing structures undergoing large deformations, conventional FEM-based TO often exhibits numerical instabilities due to severe mesh distortions, tangling, and large rotations, consequently leading to convergence failures.

To address this challenge, we present a TO framework based on the Material Point Method (MPM). MPM is a hybrid Lagrangian–Eulerian particle method, well-suited for simulating large deformations. In particular, we present an end-to-end differentiable implicit MPM framework for designing structures undergoing quasi-static hyperelastic large deformations. The effectiveness of the approach is demonstrated through validation studies encompassing both single and multi-material designs, including the design of compliant soft robotic grippers. The software accompanying this paper can be accessed at \href{https://github.com/UW-ERSL/MOTO}{github.com/UW-ERSL/MOTO}.

\end{abstract}


\input{1_introduction}

\input{1b_relatedWork}

\input{2_method}

\input{3_result}

\input{4_conclusion}


\section*{Compliance with ethical standards}
The authors declare that they have no conflict of interest.

\section*{Replication of Results}
The Python source code is available in a public GitHub repository at \href{https://github.com/UW-ERSL/MOTO}{github.com/UW-ERSL/MOTO}. This implementation is also provided as supplementary material to the version published in the journal.

\bibliographystyle{unsrt}  
\bibliography{5_references}

\end{document}

%% file: 0_preamble.tex
\usepackage{arxiv}

\usepackage[utf8]{inputenc} 
\usepackage[T1]{fontenc}    
\usepackage{hyperref}
\usepackage{url}            
\usepackage{booktabs}       
\usepackage{bm}
\usepackage{amsfonts}       
\usepackage{amsmath}
\usepackage{amssymb}
\usepackage{nicefrac}       
\usepackage{microtype}
\usepackage{cleveref}
\usepackage{mathtools}
\usepackage{xcolor}
\usepackage{algpseudocode}
\usepackage{algorithm}
\usepackage{color,soul}
\usepackage{caption}
\usepackage{lipsum}
\usepackage{fancyhdr}       
\usepackage{graphicx}       
\graphicspath{{media/}}     
\usepackage{enumitem}
\makeatletter
\newcommand{\printfnsymbol}[1]{%
  \textsuperscript{\@fnsymbol{#1}}%
}

\makeatother

%% file: 1_introduction.tex
\section{Introduction}
\label{sec:intro}

\begin{figure}[]
    \begin{center}
        \includegraphics[scale=0.34,trim={0 0 0 0},clip]{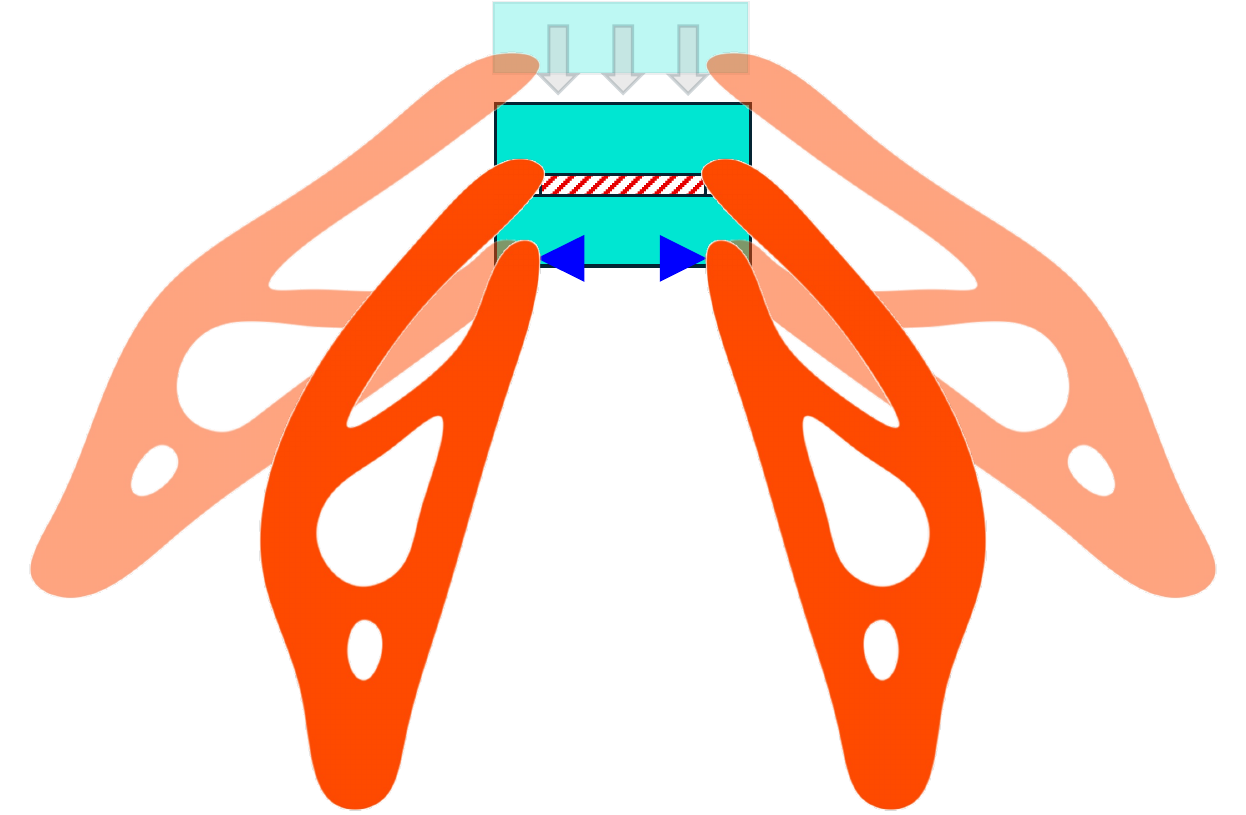}
        \caption{Design of soft robotic gripper undergoing large deformation.}
\label{fig:mpm_application}

    \end{center}
\end{figure}

Topology Optimization (TO) \cite{bendsoe2013topology, sigmund2013topology} is a computational design method that optimizes material distribution within a design domain, subject to specified objectives and constraints. Structural TO has largely focused on small deformations with linear constitutive models \cite{sigmund200199, liu2018current,wang2021comprehensive}; for a comprehensive review, we refer the reader to \cite{deaton2014survey, wein2020review}. However, modern engineering increasingly demands designs that account for large deformations and nonlinear constitutive behaviors. Prominent applications include soft robotics \cite{crowley20223d}, protective cellular structures \cite{jia2021crashworthiness, deng2020topology, giraldo2025toward}, flexible electronics \cite{luo2025nonlinear}, and pneumatic actuators \cite{kobayashi2024computational} (for example, see \Cref{fig:mpm_application}).

\begin{figure}[]
    \begin{center}
        \includegraphics[scale=0.34,trim={0 0 0 0},clip]{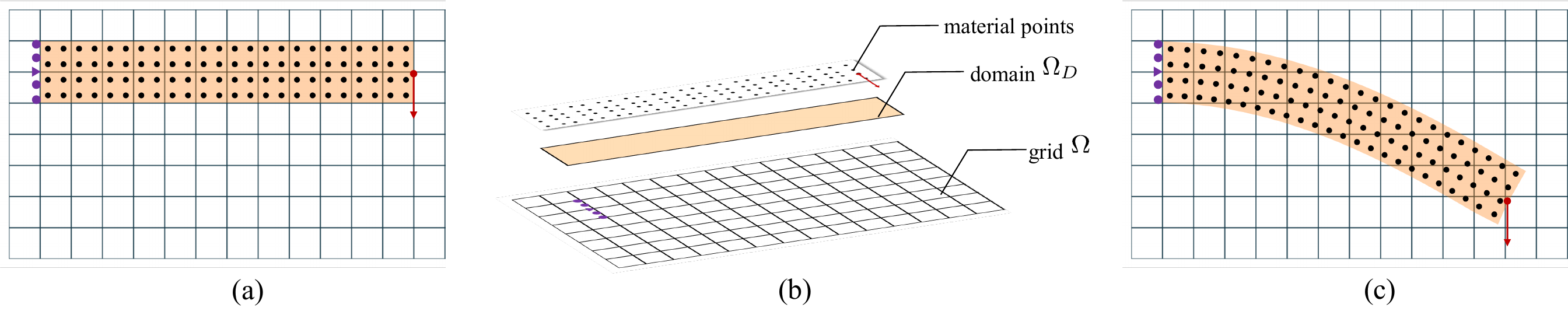}
        \caption{MPM discretization: (a) initial configuration, (b) exploded view showing material points, computational domain, and background grid, and (c) deformed configuration.}
\label{fig:method_overview}
    \end{center}
\end{figure}

Traditionally, the Finite Element Method (FEM) has been employed within TO to analyze structural responses (the forward problem) and compute sensitivities through adjoint methods (the backward problem). However, the Lagrangian mesh-based paradigm of FEM introduces significant challenges under large deformations. As elements deform with the material, they are prone to severe distortion \cite{lee1993effects}, inversion, tangling \cite{prabhune2023computationally, prabhune2024isoparametric, danczyk2013tangling}, and fragmentation \cite{cervera2011mesh, mishnaevsky2003computational}, leading to numerical instability and convergence failure. The Material Point Method (MPM) \cite{sulsky1994particle} addresses these limitations by integrating Lagrangian and Eulerian descriptions. In MPM, the continuum is represented by moving Lagrangian material points carrying physical quantities (such as mass, stress, and velocity), while an Eulerian background grid facilitates the solution of the governing equations \cite{coombs2020ample}. As illustrated in \Cref{fig:method_overview}(a), the initial body is discretized into a set of material points that reside within the computational domain. The relationship between these points, the domain, and the background grid is further clarified in the exploded view in \Cref{fig:method_overview}(b). Unlike traditional FEM, where the mesh deforms with the material, the material points in MPM move through the fixed background grid as the body undergoes deformation (\Cref{fig:method_overview}(c)). This hybrid approach effectively mitigates mesh-related issues \cite{lian2011coupling, wikeckowski2004material}.

\begin{figure}[]
    \begin{center}
        \includegraphics[scale=0.34,trim={0 0 0 0},clip]{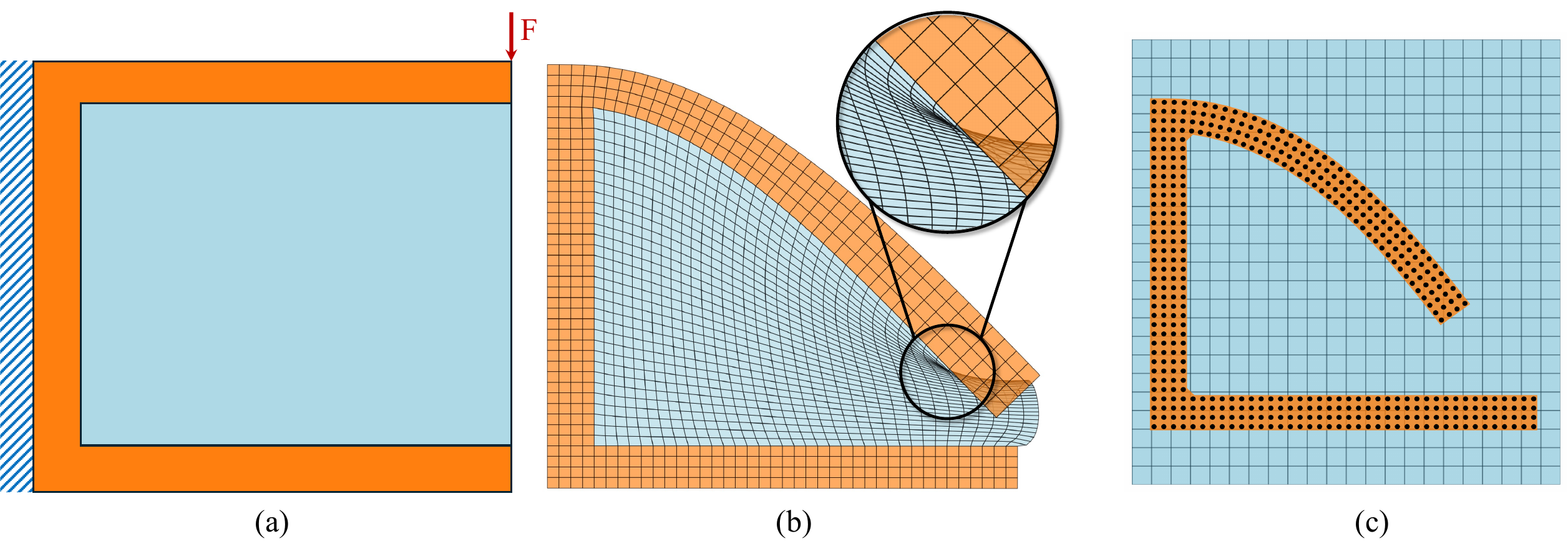}
        \caption{(a) C-bracket domain with boundary conditions. (b) Finite-element solution under the applied load, illustrating severe mesh distortion. (c) Material point method representation of the same problem, where material points move on a fixed background grid.}
        \label{fig:fem_mpm}
    \end{center}
\end{figure}

For instance, consider the C-bracket example from \cite{wang2014interpolation} shown in \Cref{fig:fem_mpm}(a). In a standard FEM formulation (\Cref{fig:fem_mpm}(b)), the mesh deforms with the material, producing highly distorted and inverted elements that result in numerical instabilities. In contrast, the MPM formulation (\Cref{fig:fem_mpm}(c)) maintains a fixed background grid while material points track the deformed configuration, effectively bypassing mesh distortion regardless of deformation complexity. This renders MPM particularly advantageous for large-deformation TO.

While MPM has been predominantly tied to explicit time integration \cite{hu2018moving} and dynamic TO applications \cite{yuhn20234d, sato2022topology}, its application to quasi-static problems remains comparatively limited. Unlike explicit methods that utilize vector-based updates, quasi-static regimes necessitate implicit solvers. This requirement introduces significant numerical challenges, as it involves the assembly and inversion of a global tangent stiffness matrix within the MPM framework. This process is further complicated in quasi-static TO by the evolving particle-grid topology.

Various implicit MPM variants, such as the Generalized Interpolation Material Point (GIMP) \cite{charlton2017igimp, coombs2020ample}, the Dual Domain Material Point (DDMP) method, and the Convected Particle Domain Interpolation (CPDI) \cite{sadeghirad2011cpdi} have been proposed. Here we adopt and extend the GIMP method as it mitigates the grid-crossing instabilities inherent in standard MPM by providing $C^1$ continuity, which is critical for the stability and convergence of implicit solvers. While this formulation was recently applied to TO for small deformation linear elasticity \cite{park2025topology}, its extension to large deformation hyperelasticity remains unexplored.

To this end, we present MOTO, an implicit $\underline{m}$aterial p$\underline{o}$int framework for $\underline{t}$opology $\underline{o}$ptimization. The primary contributions of this work are as follows:

\begin{enumerate}
    \item \textbf{Hyperelastic GIMP Formulation:} Implementation of a Hencky hyperelastic material model within the GIMP framework to capture the response of structures undergoing large deformations \cite{coombs2020ample}.

    \item \textbf{Multimaterial Design Representation:} Integration of a neural network-based design representation \cite{chandrasekhar2021multi} to enable spatially varying materials.

    \item \textbf{End-to-End Differentiability:} Utilization of Automatic Differentiation (AD) \cite{padhy2025toflux} via the JAX library \cite{jax2018github} to automate the complex sensitivity computations arising from neural representation, diverse objectives, constraints, and hyperelastic constitutive models.

    \item \textbf{Application:} A demonstration of the framework's capabilities through a case study on the design of compliant soft robotic grippers.
    
\end{enumerate}

%% file: 1b_relatedWork.tex
\section{Related Work}
\label{sec:relatedWork}
This section reviews the state of the art in TO for structures undergoing large deformations across diverse applications. In addition, we review the numerical discretization methods used in large-deformation TO, spanning mesh-based and particle-based approaches.

Since the foundational work of \cite{bendsoe1988generating}, TO has matured into a powerful computational design tool used across a wide range of engineering applications \cite{alexandersen2020review, sigmund2013topology, padhy2025photos, padhy2026tomatoes, padhy2024fluto}. Much of the classical structural TO literature has been developed under small-strain assumptions and linear constitutive models \cite{sigmund200199, liu2018current,wang2021comprehensive}. While suitable for structures undergoing small strains, these assumptions become restrictive when the structural response is governed by large deformations and nonlinear constitutive behaviors. This is the case in many contemporary applications, including soft robots undergoing large motions and hyperelastic deformation \cite{dalklint2024simultaneous}, cardiovascular stents undergoing large deployment deformations \cite{james2016layout, xue2020design}, crashworthy structures dissipating energy through collapse \cite{jia2021crashworthiness, deng2020topology, giraldo2025toward}, and architected metamaterials whose response depends on finite-strain mechanisms \cite{wang2014design, clausen2015topology, zhang2019computational, dalklint2023computational}.

Among the most active application areas of large-deformation TO are mechanical metamaterials, where the objective is often to tailor nonlinear effective response \cite{wang2014design}. \cite{clausen2015topology} demonstrated topology-optimized architectures with programmable Poisson’s ratios for large deformation. Building on this direction, \cite{zhang2019computational} coupled density-based TO with nonlinear homogenization to design finite-strain auxetic metamaterials. More recently, \cite{li2021design} used multimaterial TO to obtain composite metastructures with programmable force–displacement responses. Another important application area of large deformation TO is biomedical design, where performance often depends on large deployment, expansion, or compliant motion under physiological loading. \cite{james2016layout} designed a bi-stable cardiovascular stent whose deployment relies on snap-through between contracted and expanded stable configurations. 

An important class of large-deformation TO problems involves contact, where the contact interface is generally unknown a priori and evolves together with the topology. \cite{fernandez2020topology} developed a framework for TO of multiple deformable bodies in contact using large-deformation mechanics. \cite{bluhm2021internal} extended the third-medium contact method to finite-strain TO with self-contact, introducing a regularization of the void region that makes the method suitable for very large deformations. Another practical application of large deformation TO is soft grippers, as they rely on material compliance to interact safely with delicate objects and adapt to diverse geometries \cite{crowley20223d, rus2015design}. TO of a cable-driven soft gripper with geometric nonlinearity was presented in \cite{wang2020topology}. Furthermore,  a framework for pressure-driven soft robots using nonlinear TO was proposed in \cite{caasenbrood2020computational}. Simultaneous shape and topology optimization of inflatable soft robots using finite-deformation hyperelasticity was formulated in \cite{dalklint2024simultaneous}. Finally, large-deformation TO has also become important in crashworthiness \cite{jia2021crashworthiness}, impact mitigation \cite{deng2020topology, giraldo2025toward}, and elastoplastic design \cite{jia2026multimaterial}, where performance depends on irreversible deformation and energy dissipation. \cite{wallin2016topology} extended elastoplastic TO to finite strains using rate-independent isotropic hardening plasticity and a multiplicative split of the deformation gradient. More recently, \cite{jia2026multimaterial} presented a multimaterial finite-strain elastoplastic formulation that simultaneously optimizes structural layout and material distribution, thereby expanding the design space for energy-dissipating and forming-dominated applications.

Underlying these diverse applications is a common computational challenge, namely the development of numerical discretization methods that can robustly handle large deformations during TO. Historically, most TO frameworks have relied on Lagrangian finite element discretizations \cite{sigmund200199}, which provide accurate structural analysis in the small-deformation, linear-elastic regime and support density-based \cite{andreassen2011efficient}, level-set \cite{allaire2004structural}, and phase-field based TO formulations \cite{takezawa2010shape}. Extensions to large deformations have commonly been pursued through updated Lagrangian finite element formulations \cite{buhl2000stiffness}. However, in density-based TO under large deformation, low-density regions can undergo severe mesh distortion, which may lead to element inversion and ill-conditioned stiffness matrices \cite{lahuerta2013towards,lee1993effects}. Several stabilization strategies have therefore been proposed within nonlinear FEM, including element removal and reintroduction schemes \cite{bruns2003element,yoon2005element}, energy-interpolation approaches for finite-strain fictitious-domain formulations \cite{wang2014interpolation}, and additive hyperelastic regularization of distortion-prone elements \cite{luo2015topology}. Alternative discretizations have also been explored. Meshfree methods such as the element-free Galerkin method and reproducing kernel particle methods remove persistent mesh connectivity \cite{belytschko1994element,neofytou2020level} and have been applied in TO \cite{shobeiri2015topology,zheng2015topology}. However, they introduce additional complexity in enforcing boundary conditions and maintaining accuracy \cite{garg2018meshfree,krongauz1996enforcement}.

Particle-based methods offer an alternative route for handling large motion and evolving geometry. In particular, the Material Point Method (MPM) combines Lagrangian material points with an Eulerian background grid and is well suited to large deformation and contact \cite{sulsky1994particle,de2020material}. Beyond its established use in geomechanics \cite{soga2016trends} and impact simulations \cite{huang2011contact}, MPM has recently been integrated with TO \cite{park2025topology}, where the authors developed a quasi-static MPM-based TO framework in the small-deformation, linear-elastic regime, including analytical sensitivities and a study of cell-crossing effects. For dynamic problems, MPM-based TO has been used for trajectory-dependent objectives in which inertia is intrinsic to the formulation \cite{yuhn20234d, sato2023topology}. In contrast, many design problems are governed by sequences of quasi-static equilibrium states, e.g., controlled actuation or sustained loading while maintaining stability \cite{shiva2019elasticity,zhou2022modeling,lee2023quasi}. For such applications, \emph{MPM-based TO for quasi-static problems involving large deformation remains comparatively underexplored, motivating the present work}.

%% file: 2_method.tex
\section{Proposed Method}
\label{sec:method}

This work introduces a topology optimization (TO) framework based on the Generalized Interpolation Material Point (GIMP) method \cite{coombs2020ample}, which is tailored for structures undergoing large deformations. We begin by prescribing a grid $\Omega$, within which the design domain $\Omega_D$ is populated with material points as illustrated in \Cref{fig:method_overview}(b). The grid $\Omega$ is chosen sufficiently large to accommodate all expected deformed configurations of the body throughout the analysis. We assume that Dirichlet boundary conditions are applied to the grid nodes, while external loads are applied directly to the material points. We also assume the material constants for the assumed constitutive model (Hencky hyperelastic \cite{coombs2020ample} in this work) have been prescribed. The primary objective is to determine the optimal material distribution that maximizes performance subject to design constraints.

The remainder of the methodology to achieve this objective is organized as follows: the governing equations are detailed in \Cref{sec:method_govEq}. The TO formulation is presented in \Cref{sec:method_topopt}. Sensitivity analysis via automatic differentiation is discussed in \Cref{sec:method_sensAnalysis}, and the complete optimization algorithm is summarized in \Cref{sec:method_algorithm}.

 For the sake of completeness, we begin by reviewing the material point method in \Cref{sec:method_mpmOverview}. For a detailed treatment, we refer the readers to \cite{coombs2020ample}.
 
\subsection{Overview of Material Point Method}
\label{sec:method_mpmOverview}

\begin{figure}[]
    \begin{center}
\includegraphics[scale=0.35,trim={0 0 0 0},clip]{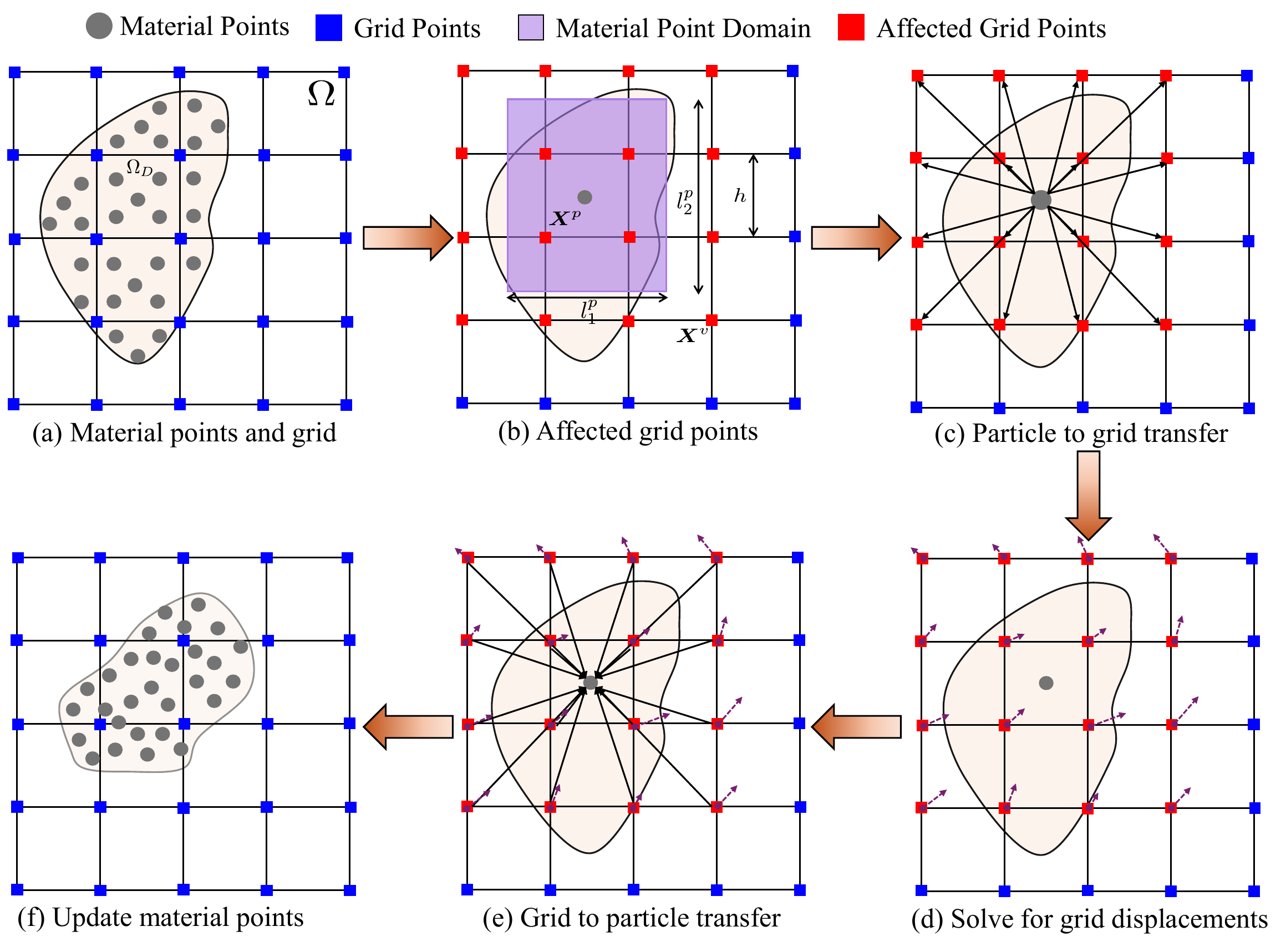}
        \caption{Illustration of the GIMP-based particle-grid coupling and update procedure. Material points are shown as gray circles, grid points as blue squares, and the domain of a representative material point as the shaded region. Grid points with nonzero GIMP support for that material point are highlighted in red. Panels (a)--(f) summarize the main MPM steps: initialization, identification of affected grid points, particle-to-grid transfer, solution of the force balance equations, grid-to-particle transfer, and update of the material points.}
        \label{fig:mpm_discretization}
    \end{center}
\end{figure}

\begin{figure}[]
    \begin{center}
\includegraphics[scale=0.75,trim={0 0 0 0},clip]{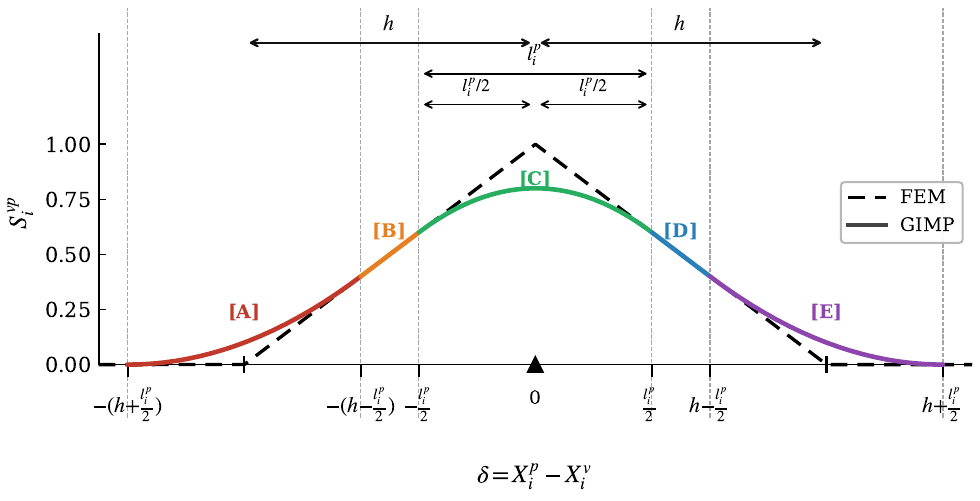}
        \caption{One-dimensional GIMP basis function \(S_i^{vp}\) versus the relative coordinate \(\delta = X_i^p - X_i^v\), compared with the conventional finite-element basis function. Regions [A]--[E] correspond to the five cases in \Cref{eq:Svp}.}
        \label{fig:gimp_shape_function}
    \end{center}
\end{figure}

We detail the numerical procedure of the Material Point Method (MPM) in this section. MPM combines Lagrangian and Eulerian descriptions to analyze structures undergoing large deformations. In this hybrid formulation, the continuum body is discretized into a set of Lagrangian material points (MPs) that are associated with the physical state variables, including mass, volume, deformation gradient, and stress. The governing force balance equations are solved on a fixed Eulerian grid.  For instance, consider \Cref{fig:mpm_discretization} which details the core MPM steps as follows:

\begin{enumerate}
    \item \textbf{Initialization:} The design domain $\Omega_D$ is populated with $n_p$ material points, each assigned a uniform initial volume $V_p^0$, mass $m_p = \rho V^0_p$, and identity deformation gradient. These points reside within a computational domain $\Omega$ composed of $n_n$ grid nodes as illustrated in \Cref{fig:mpm_discretization}(a). Furthermore, the Dirichlet boundary conditions are imposed on the grid nodes, while the loads are applied to the material points.
    \item \textbf{Material point influence:}   Each material point $p$ interacts with the set of grid nodes whose GIMP basis functions have nonzero support over its influence domain, defined by lengths $\bm{l}^p$, as shown in \Cref{fig:mpm_discretization}(b). The mapping between the material points and the grid nodes is governed by the product-form GIMP shape functions $S^{vp} = \prod\limits_i S^{vp}_i$, where the one-dimensional factors  are defined in \Cref{eq:Svp}. 

\begin{equation}
\label{eq:Svp}
  S_i^{vp} =
  \begin{cases}
    \dfrac{\bigl(h + \tfrac{l_i^p}{2} + X_i^p - X_i^v\bigr)^{2}}
          {2\,h\,l_i^p},
    & -h - \tfrac{l_i^p}{2} < X_i^p - X_i^v \leq -h + \tfrac{l_i^p}{2},
      \quad \textup{[A]} \\[8pt]
    1 + \dfrac{X_i^p - X_i^v}{h},
    & -h + \tfrac{l_i^p}{2} < X_i^p - X_i^v \leq -\tfrac{l_i^p}{2},
      \quad \textup{[B]} \\[8pt]
    1 - \dfrac{(X_i^p - X_i^v)^{2} + \bigl(\tfrac{l_i^p}{2}\bigr)^{2}}
              {h\,l_i^p},
    & -\tfrac{l_i^p}{2} < X_i^p - X_i^v \leq \tfrac{l_i^p}{2},
      \quad \textup{[C]} \\[8pt]
    1 - \dfrac{X_i^p - X_i^v}{h},
    & \tfrac{l_i^p}{2} < X_i^p - X_i^v \leq h - \tfrac{l_i^p}{2},
      \quad \textup{[D]} \\[8pt]
    \dfrac{\bigl(h + \tfrac{l_i^p}{2} - X_i^p + X_i^v\bigr)^{2}}
          {2\,h\,l_i^p},
    & h - \tfrac{l_i^p}{2} < X_i^p - X_i^v \leq h + \tfrac{l_i^p}{2}.
      \quad \textup{[E]}
  \end{cases}
\end{equation}

Here $h$ denotes the grid cell size, $X^p_i$ is the material point coordinate, and $X^v_i$ is the node coordinate along dimension $i$. The spatial gradient $\nabla_x S^{vp}$ is obtained analytically from the above by differentiating with respect to the current configuration coordinates \cite{coombs2020ample}.

The one-dimensional basis functions are illustrated in \Cref{fig:gimp_shape_function}, where the horizontal axis denotes the relative position of the material point $\delta = X_i^p - X_i^v$. The regions [A]--[E] shown in the figure correspond directly to the five cases in \Cref{eq:Svp}. In regions [B] and [D], the generalized interpolation basis functions coincide with the conventional finite-element basis functions as the material point domain lies within a single grid cell. In regions [A], [C], and [E], by contrast, the material point domain spans multiple background grid cells, and the basis functions therefore differ from the conventional finite-element form.

    \item \textbf{Particle to grid transfer:} Once the grid nodes influenced by each material point have been identified, the force contributions are mapped from the material points to the grid as illustrated in \Cref{fig:mpm_discretization}(c). Specifically, the external nodal force is assembled from body forces and prescribed particle loads using $S^{vp}$, whereas the internal nodal force is assembled from the particle stress state using the shape-function gradients, as described in \Cref{sec:method_govEq}.
    
    \item \textbf{Solve on the grid:} 
    With the nodal force vectors determined and the boundary conditions imposed, we solve the force balance equations on the grid to obtain the nodal displacement $ \mathbf{u}_v$ (see \Cref{fig:mpm_discretization}(d)). As the problem is nonlinear, this force balance equations are solved iteratively using a Newton-Raphson procedure  (see \Cref{sec:method_govEq} for details).

    \item \textbf{Grid to particle transfer:} After the grid solution is obtained, the nodal displacement field is interpolated back to the material points using the same GIMP shape functions $S^{vp}$, thereby yielding the displacement of each material point as illustrated in \Cref{fig:mpm_discretization}(e).
    
    \item \textbf{Update material points:} From the updated material point displacements, all physical state variables are updated: the deformation gradient, the Cauchy stress, the material point volume, GIMP domain half-lengths, and the material point positions as shown in \Cref{fig:mpm_discretization}(f). Critically, the quantities stored on the grid are reset, whereas the updated state variables carried by the material points are retained. For highly nonlinear problems, the total applied load is divided into increments, and the above steps are repeated for each increment.
\end{enumerate}

\subsection{Governing Equations and Solution Procedure}
\label{sec:method_govEq}

Having described the numerical procedure of MPM, we now discuss the governing equations.  In particular, to capture the structural response in the large deformation regime, we adopt a framework that accounts for geometric nonlinearity. Additionally, we employ a hyperelastic material model to describe the constitutive behavior.

Consider a design occupying a computational domain $\Omega$ as shown in \Cref{fig:method_overview}. The quasi-static response is governed by the force balance equation: \Cref{eq:strong_form}:

\begin{equation}
\label{eq:strong_form}
R(\mathbf{u}) \coloneqq
  \nabla \cdot \boldsymbol{\sigma} + \rho \mathbf{b} = \mathbf{0} \quad,\;  \text{for } \mathbf{x} \in \Omega
\end{equation}

\noindent where $\mathbf{u}$ is the displacement field, $\boldsymbol{\sigma}$ is the Cauchy stress tensor, $\rho$ is the material density, and $\mathbf{b}$ is the specific body force. The prescribed displacement $\bar{\mathbf{u}}$ is applied on the grid nodes. Furthermore, point loads are applied on the material points at the prescribed locations. In the present work, traction boundary conditions are not considered. Their imposition in the MPM is nontrivial, as it generally requires an explicit representation of the physical boundary \cite{bing2019b}. We therefore adopt a simplified treatment in which Dirichlet boundary conditions and point loads are considered.

We adopt an updated Lagrangian framework to describe the large deformation kinematics \cite{coombs2020ample}. The deformation is characterized by the deformation gradient $\mathbf{F}$, defined as:

\begin{equation} 
\label{eq:deformation_gradient} F_{ij} = \frac{\partial x_i}{\partial X_j} 
\end{equation}

\noindent \noindent where $X_i$ and $x_i$ denote the coordinates of material point $p$ at the beginning of the load step and in the deformed configuration, respectively. For the constitutive response, we employ a Hencky hyperelastic model. The elastic logarithmic (Hencky) strain $\boldsymbol{\varepsilon}^e$ is defined as:
\begin{equation}
\label{eq:log_strain}
\boldsymbol{\varepsilon}^e = \frac{1}{2} \ln(\mathbf{B}), \quad \text{where} \quad \mathbf{B} = \mathbf{F} \mathbf{F}^\top
\end{equation}
\noindent is the left Cauchy- Green strain tensor and $\ln(\cdot)$ denotes the logarithm to base $e$. The Kirchhoff stress $\boldsymbol{\tau}$ is obtained using the isotropic linear elastic relation:

\begin{equation}
\label{eq:kirchhoff_stress}
\boldsymbol{\tau}
= \lambda\,\mathrm{tr}(\boldsymbol{\varepsilon}^e)\,\mathbf{I}
+ 2\mu\,\boldsymbol{\varepsilon}^e .
\end{equation}

\noindent where $\lambda$ and $\mu$ are the Lam\'{e} parameters. Furthermore, for structures with small thickness relative to the in-plane dimensions, we enforce a plane-stress condition ($\sigma_{zz}=0$), which yields:
\begin{equation}
\label{eq:plane_stress_closure}
\varepsilon_{zz}^e = -\frac{\lambda}{\lambda + 2\mu} (\varepsilon_{xx}^e + \varepsilon_{yy}^e)
\end{equation}

Finally, the Cauchy stress is subsequently recovered via $\boldsymbol{\sigma} = \frac{1}{J} \boldsymbol{\tau}$, where $J = \det(\mathbf{F})$ \cite{Neto2008computationalMechanics}.

Building upon the established MPM numerical procedure (\Cref{sec:method_mpmOverview}), the discrete quasi-static force balance is enforced on the grid by assembling nodal internal and external force vectors. The internal force vector $\mathbf{f}^{\mathrm{int}}_v$ is assembled from the particle
stress contributions, which are defined as:
\begin{equation}
\label{eq:f_int_particle}
\mathbf{f}^{\mathrm{int}}_{v,p}
= V_p\,(\nabla_x S^{vp})^\top\,\boldsymbol{\sigma}_p ,
\end{equation}
where $\nabla_x S^{vp}$ denotes the spatial gradient of the GIMP shape function (\Cref{eq:Svp}), $\boldsymbol{\sigma}_p$ is the Cauchy stress at material point $p$, and $V_p$ is the material point volume in the current configuration.

The external force vector $\mathbf{f}^{\mathrm{ext}}_v$ is assembled from particle contributions due to body forces and applied point loads, which are defined as follows:
\begin{equation}
\label{eq:f_ext_particle}
\mathbf{f}^{\mathrm{ext}}_{v,p}
= S^{vp}\, m_p\, \mathbf{b}_p + S^{vp}\, \mathbf{f}^{\mathrm{ext}}_p ,
\end{equation}
where $m_p=\rho V_p$ is the material point mass, $\mathbf{b}_p$ is the specific body force evaluated at material point $p$, and $\mathbf{f}^{\mathrm{ext}}_p$ denotes externally applied point forces at material point $p$.

Combining \Cref{eq:f_int_particle} and \Cref{eq:f_ext_particle}, the discretized residual equation is:
\begin{equation}
\label{eq:residual_global}
\mathbf{R}( \mathbf{u})
 = \mathbf{f}^{\mathrm{int}}(\mathbf{u}) - \mathbf{f}^{\mathrm{ext}} = \mathbf{0}
\end{equation}
Here, $\mathbf{f}^{\mathrm{int}}$ and $\mathbf{f}^{\mathrm{ext}}$ denote the global nodal force vectors obtained by assembling the nodal contributions $\mathbf{f}^{\mathrm{int}}_v$ $\mathbf{f}^{\mathrm{ext}}_v$ for all grid nodes $v$.

Due to geometric and material nonlinearity, \Cref{eq:residual_global} is solved using a Newton-Raphson scheme.
At iteration $k$, the displacement correction $\delta \mathbf{u}^{(k)}$ is obtained from
\begin{equation}
\label{eq:newton_system}
\mathbf{K}^{(k)}\,\delta \mathbf{u}^{(k)} = -\mathbf{R}( \mathbf{u}^{(k)}),
\end{equation}

The displacement is then updated $ \mathbf{u}^{(k+1)} = \mathbf{u}^{(k)} + \delta \mathbf{u}^{(k)}$ until convergence. The global stiffness matrix $\mathbf{K}$ is defined as:

\begin{equation}
\label{eq:k_global}
\mathbf{K}^{(k)}
=
\left.\frac{\partial \mathbf{R}}{\partial \mathbf{u}}\right|_{\mathbf{u}^{(k)}}
\end{equation}

Rather than deriving the tangent stiffness matrix analytically for the finite-strain kinematics and GIMP mapping, we compute it via automatic differentiation within JAX \cite{jax2018github}. Furthermore, this end-to-end differentiability enables the sensitivity analysis required for TO, as discussed in \Cref{sec:method_sensAnalysis}.

\subsection{Solver validation}
\label{sec:method_solverValidation}

Before detailing the optimization procedure, we validate the method's accuracy and its implementation. We compare the accuracy of our implicit MPM solver against a classical benchmark: large-deformation bending of an elastic cantilever beam. This problem involves finite rotations and large strains comparable to those encountered in the TO examples. Furthermore, the problem has an analytical solution. 

\begin{figure}[h!]
    \begin{center}
\includegraphics[scale=0.31,trim={0 0 0 0},clip]{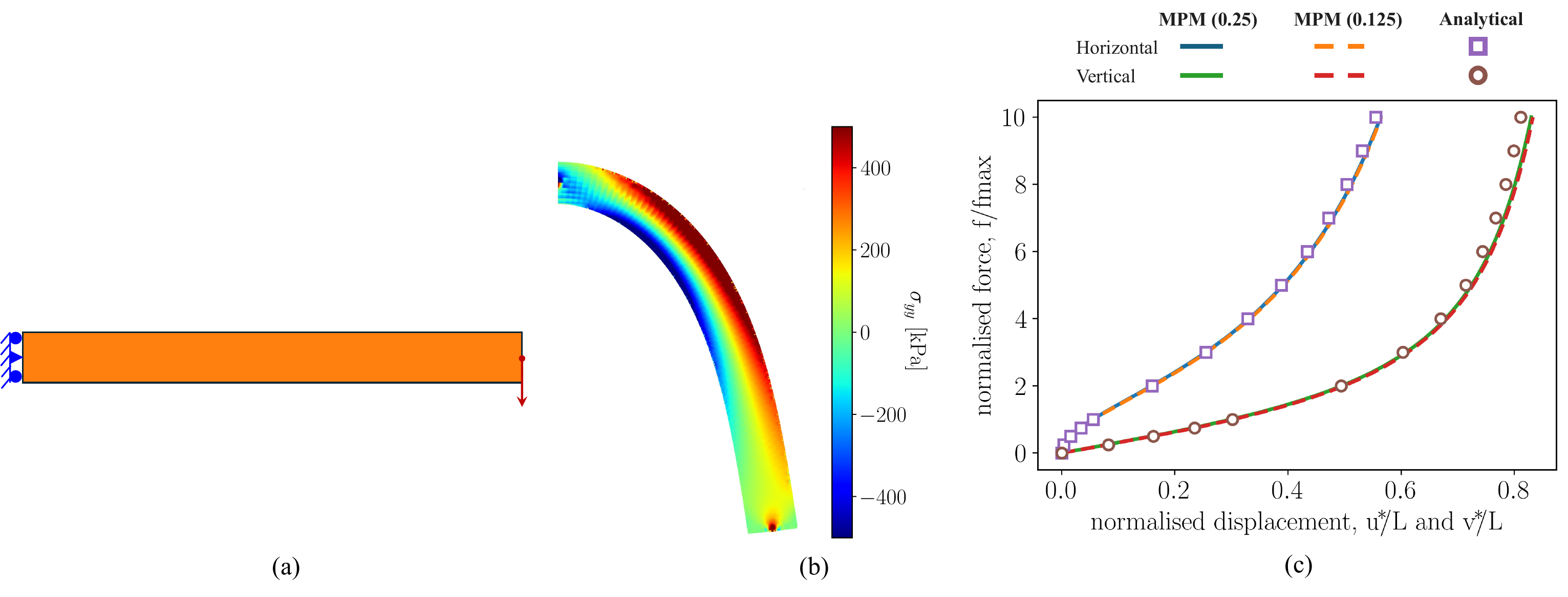}
        \caption{Validation of our implicit MPM framework for the large-deformation bending of a cantilever beam. (a) Beam geometry, boundary conditions, and loading. (b) Deformed configuration at the final load step for the $h=0.125$ m discretization, colored by the Cauchy stress component $\sigma_{yy}$. (c) Normalized force-displacement response comparing MPM predictions for two mesh resolutions with the analytical solution of \cite{molstad1977finite}.}
        \label{fig:cant_val}
    \end{center}
\end{figure}

The benchmark consists of a cantilever beam of length $L = 10$ m and width $d_0 = 1$ m, as illustrated in \Cref{fig:cant_val}(a). Sliding boundary conditions are applied along the left edge, and the beam is fully fixed at its midpoint. A concentrated vertical force is applied at the midpoint of the right edge. The beam undergoes large deflections, with tip displacements approaching the beam length. The analytical solution for this problem was derived by \cite{molstad1977finite}.

For validation, the beam is modeled with the following material properties for the Hencky hyperelastic model: Young's modulus $E = 12$ MPa and Poisson's ratio $\nu = 0.2$. A total force of $f_0 = 100$ kN is applied incrementally over 50 equal load steps. The domain is discretized using the generalized interpolation material point method (GIMP) with 36 material points per background grid cell. Two mesh resolutions are tested: $h = 0.25$ m and $h = 0.125$ m.

\Cref{fig:cant_val}(b) shows the deformed configuration at the final load step for the $h = 0.125$ m discretization. The beam undergoes substantial bending, with the tip rotating nearly 90 degrees. The color contour displays the Cauchy stress component $\sigma_{yy}$, which exhibits the expected bending distribution: tension on the outer region and compression on the inner region.

\Cref{fig:cant_val}(c) compares the normalized force-displacement response against the analytical solution. Both horizontal ($u^*/L$) and vertical ($v^*/L$) tip displacements are plotted against the normalized applied force ($f/f_{\max}$). The MPM results for both mesh resolutions show excellent agreement with the analytical solution across the entire loading range. The difference between the two discretizations is negligible, indicating mesh convergence. At the maximum load, the vertical tip displacement reaches approximately $0.8L$, demonstrating the framework's capability to handle large deformations accurately.

\subsection{Topology Optimization} 
\label{sec:method_topopt}

With the established and validated GIMP-based MPM framework, we now formulate the TO problem. The goal is to find the optimal material distribution within $\Omega_D$ the specified objective and constraints. In the present work, we consider both single- and multi-material TO. We begin with a single-material formulation based on particle-wise pseudodensity, which serves as the most direct extension of standard density-based TO to the MPM setting. We then introduce a more general multi-material formulation based on a coordinate-based neural network. The following sections detail the proposed methodology.

\subsubsection{Single-Material Topology Optimization}
\label{sec:method_topopt_single_material}
For single-material TO, we adopt a particle-wise pseudodensity representation. This formulation is a natural extension of the standard element-wise density approach used in finite-element-based TO to the material point setting.

Specifically, we assign each material point $p$ a pseudodensity $\gamma_p \in [0,1]$, where $\gamma_p=1$ denotes solid material and $\gamma_p=0$ denotes void. Intermediate values are allowed during optimization to facilitate gradient-based optimization. Observe that in contrast to conventional FEM formulations, where the design variables are associated with grid elements, here the design variables are carried directly by the material points. Consequently, the material distribution is transported with the Lagrangian particles throughout the updated-Lagrangian MPM solve.

\subsubsection{Multi-Material Topology Optimization}
\label{sec:method_topopt_multi_material}

Many applications require more than a single material. For instance, soft robotic grippers must balance flexibility with structural rigidity, requiring design with spatially varying stiffness. While the particle-wise pseudodensity formulation could, in principle, be extended to multi-material TO, we instead adopt a coordinate-based neural-network representation. This representation subsumes the single-material case as a special case but is introduced here for multi-material design due to advantages discussed next.

Multi-material design requires specifying what material exists at each location. Towards this, let $S$ denote the number of a priori prescribed materials. Then at each point $\bm{x}$ we define volume fractions $\bm{v}(\bm{x}) = [v_1, \dots, v_S]$, where  $\sum\limits_{s=1}^{S} v_s(\bm{x}) = 1$ and $v_s(\bm{x}) \geq 0 \; ; \; \forall \bm{x}$.

\begin{figure}[]
    \begin{center}
\includegraphics[scale=0.65,trim={0 0 0 0},clip]{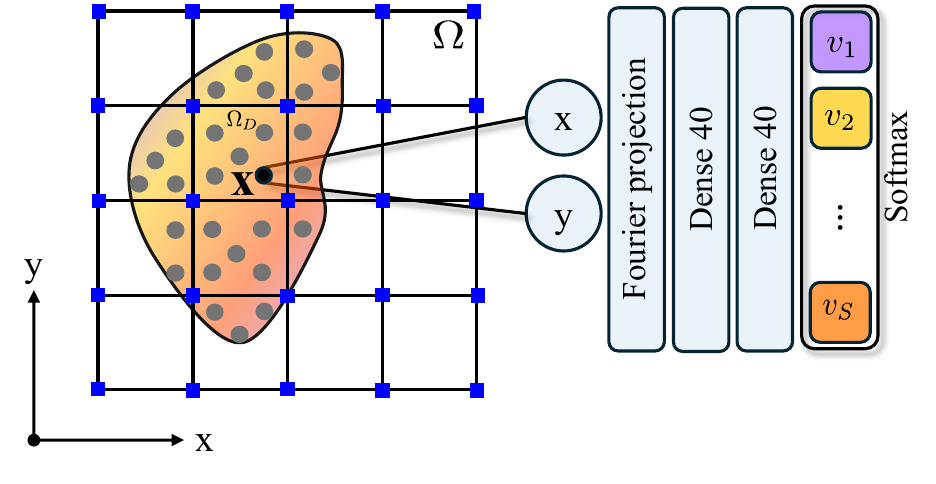}
        \caption{Coordinate-based neural network used for multi-material TO. The network maps spatial coordinates in the design domain to material volume fractions.}
        \label{fig:method_network}
    \end{center}
\end{figure}

Adopting \cite{chandrasekhar2021multi}, we parameterize this field using a neural network (NN) $\bm{v}(\bm{x}; \bm{w})$, where $\bm{w}$ denotes the weights of the NN. The NN maps spatial coordinates from the design domain to the volume fractions  \Cref{fig:method_network}. This representation offers three key advantages over conventional approaches where design variables are explicitly tied to the simulation discretization (in this case, the material points):
\begin{itemize}
    \item The network architecture inherently enforces the partition-of-unity constraint via its output activation.
    \item The design is decoupled from the background grid, allowing for a compact set of design variables independent of simulation resolution. 
    \item The material field is analytically defined everywhere, facilitating the recovery of crisp, high-resolution topologies.
\end{itemize}

The proposed network architecture, as illustrated in \Cref{fig:method_network} consists of the following components:
\begin{enumerate}
    \item \textbf{Input Layer:} The spatial coordinates $\bm{x} \in \mathbb{R}^d$ ($d = 2$ dimensions) are propagated through the network as input.
    
    \item \textbf{Fourier Projection:} The Euclidean coordinates are mapped to a higher-dimensional feature space using a Fourier projection layer to mitigate the spectral bias of standard multi-layer perceptrons and enable the representation of high-frequency topological features  \cite{tancik2020fourier}. This projection promotes faster convergence and allows the network to capture intricate structural details that would otherwise be smoothed out.

    \item \textbf{Hidden Layers:} The feature vector is propagated through a series of dense hidden layers (two layers with 40 neurons each) using ReLU activation functions \cite{agarap2018deep}.
    
    \item \textbf{Output Layer:} The final layer consists of $S$ neurons corresponding to the volume fractions of the $S$ materials. A Softmax activation function is applied to the output, intrinsically satisfying $\sum\limits_{s=1}^S v_s = 1$ and $0 \le v_s \le 1$.
\end{enumerate}

The trainable weights and biases $\bm{w}$ of this network constitute the primary design variables for the optimization process. Furthermore, the weights $\bm{w}$ are initialized using Xavier normal initialization \cite{glorot2010understanding}. The network is initialized to produce a spatially uniform volume-fraction field across all phases.

\subsubsection{Material Model}
\label{sec:method_simp}

Having described the design representation, we now detail the interpolation of the material properties. In particular, we use the standard Solid Isotropic Material with Penalization (SIMP) \cite{bendsoe1999material} scheme for the Lame parameters. Given the Lame parameters $ \lambda_1, \ldots, \lambda_S,  \mu_1, \ldots, \mu_S$ of the $S$ available materials, the effective Lame parameters $\lambda_p, \mu_p$ can be obtained as:

\begin{equation}
\lambda_p = \sum_{s=1}^{S} (v_{s,p})^{q} \lambda_s, \quad \mu_p = \sum_{s=1}^{S} (v_{s,p})^{q} \mu_s
\label{eq:simp_interp}
\end{equation}
 Furthermore, we employ standard continuation schemes \cite{sigmund1998numerical} for the penalty parameter $q$. For a single-material design, \Cref{eq:simp_interp} reduces to:

\begin{equation}
\lambda_p = \gamma_p^{q} \lambda_0, \quad \mu_p = \gamma_p^{q} \mu_0
\label{eq:simp_single}
\end{equation}

Furthermore, the mass density is interpolated linearly for multi-material formulation as:
\begin{equation}
\rho_p = \sum_{s=1}^{S} v_{s,p} \rho_s
\label{eq:density_interp}
\end{equation}
For single-material formulation, this reduces to $\rho_p = \gamma_p \rho$.
\subsubsection{Design Objective}
\label{sec:method_objectives}

Having obtained the interpolated material properties, we now discuss the design objective. We consider two formulations in our examples, namely, the minimization of compliance and the design of compliant mechanisms.

\paragraph{Compliance:}

We consider the design of load-bearing structures by minimizing the structural compliance:
\begin{equation}
J_c = \mathbf{f}^{\mathrm{ext}\top} \mathbf{u^*}
\label{eq:compliance}
\end{equation}
where $\mathbf{f}^{\mathrm{ext}}$ is the external force and $\mathbf{u^*}$ is the converged material point displacement.

\paragraph{Compliant Mechanisms:}

We also consider the design of compliant soft robotic grippers. Here, the goal is to maximize motion transmission from input to output. In other words, an input actuation, such as a pneumatic pressure, must produce a desired gripping motion at the output port. Concurrently, the mechanism must maintain sufficient rigidity to grasp objects without excessive internal deformation. We quantify this trade-off through the ratio of mutual strain energy ($\text{MSE}$) to the sum of input ($\text{SE}_{\text{in}}$) and output ($\text{SE}_{\text{out}}$) strain energies \cite{zhu2020designCompliantMechanism}:

\begin{equation}
J_m = -\frac{\mathrm{MSE}}{\mathrm{SE_{in}} + \mathrm{SE_{out}}} = -\frac{\mathbf{f}^{\mathrm{ext}\top}_{in}\mathbf{v^*}}{\mathbf{f}^{\mathrm{ext}\top}_{in}\mathbf{u^*} + \mathbf{f}^{\mathrm{ext}\top}_{out}\mathbf{v^*}}
\label{eq:mechanism}
\end{equation}

Here $\mathbf{u^*}$ is the displacement field under the actual input load $\mathbf{f}^{\mathrm{ext}}_{in}$, and $\mathbf{v^*}$ is the displacement field under a pseudo load $\mathbf{f}^{\mathrm{ext}}_{out}$ applied at the output port in the desired actuation direction. 

\subsubsection{Constraints}
\label{sec:method_constraints}
To ensure cost-effective designs, we impose constraints on the amount of material used.

\paragraph{Volume Constraint:}
For single-material design, we impose the standard volume constraint as:
\begin{equation}
g_v \equiv \frac{1}{V^*}\sum_{p=1}^{n_p} V_p^0 \gamma_p - 1 \leq 0
\label{eq:vol_constraint}
\end{equation}

where $V_p^0$ is the reference volume of particle $p$ and $V^*$ is the prescribed volume limit.

\paragraph{Mass Constraint:}
For multi-material design, we  enforce a mass constraint:

\begin{equation}
g_m \equiv \frac{1}{M^*}\sum_{p=1}^{n_p} V_p^0 \rho_p(\bm{v}_p) - 1 \leq 0
\label{eq:mass_constraint}
\end{equation}

where $\rho_p$ is the interpolated density from \Cref{eq:density_interp} and $M^*$ is the allowable mass.


\subsection{Sensitivity Analysis}
\label{sec:method_sensAnalysis}

A critical step in gradient-based TO is the computation of sensitivities, the derivatives of objectives and constraints with respect to design variables. The sensitivity analysis for this work is particularly complex because it must account for nonlinear hyperelasticity with large-strain kinematics and a computational pipeline that integrates the MPM solver with a coordinate-based neural network.

To address this complexity, we construct an end-to-end differentiable pipeline using the automatic differentiation (AD) capabilities of the JAX framework. This allows us to avoid the laborious and error-prone process of manually deriving sensitivity expressions. By implementing the entire forward analysis within JAX, from the neural design mapping to the nonlinear MPM solver, the framework computes the required derivatives to machine precision using reverse-mode AD.

We emphasize three challenges inherent to our structural simulation. The first arises from the use of iterative schemes, such as the Newton-Raphson method, to solve the nonlinear governing equations at each load step. A naive application of AD would unroll derivative computation across every solver iteration, computing the total derivative via the chain rule:
\begin{equation}
\frac{\mathrm{d} \mathbf{u}^{(K)}}{\mathrm{d} \boldsymbol{\gamma}} = \frac{\partial \mathbf{u}^{(K)}}{\partial \mathbf{u}^{(K-1)}} \cdot \frac{\partial \mathbf{u}^{(K-1)}}{\partial \mathbf{u}^{(K-2)}} \cdot \ldots \cdot \frac{\partial \mathbf{u}^{(1)}}{\partial \mathbf{u}^{(0)}} \cdot \frac{\partial \mathbf{u}^{(0)}}{\partial \boldsymbol{\gamma}}
\label{eq:loop_unroll}
\end{equation}
This approach is computationally expensive and memory-intensive, as both scale with the number of iterations $K$. To overcome this inefficiency, we apply the Implicit Function Theorem (IFT) \cite{blondel2022ImplicitDifferentiation}. Given that the residual vanishes at convergence, $\mathbf{R}(\boldsymbol{\gamma}, \mathbf{u}^{(K)}) = \mathbf{0}$, and the tangent stiffness $\mathbf{K} = \partial \mathbf{R} / \partial \mathbf{u}$ is invertible, the derivative follows directly:
\begin{equation}
\frac{\mathrm{d} \mathbf{u}^{(K)}}{\mathrm{d} \boldsymbol{\gamma}} = -\mathbf{K}^{-1} \frac{\partial \mathbf{R}}{\partial \boldsymbol{\gamma}}
\label{eq:ift}
\end{equation}
This enables direct computation of derivatives from the final converged solution, bypassing the need to backpropagate through the iterative solution history.

The second challenge stems from the incremental nature of the simulation, which creates a significant memory bottleneck for the adjoint sensitivity method. Computing gradients for history-dependent objectives requires access to the structural state from all previous load steps, and storing this entire configuration history is often infeasible for long loading sequences. We mitigate this issue by employing a checkpointing scheme \cite{wang2009minimal, james2015topology}. This technique stores the system's state at select increments, and during the backward pass, intermediate configurations are recomputed on the fly. This method reduces memory requirements at the cost of a moderate increase in computation time, making the optimization of large-scale nonlinear problems feasible.

The third challenge arises from differentiating through the hyperelastic constitutive model. The Hencky strain (\Cref{eq:log_strain}) requires the matrix logarithm of $\mathbf{B}$. Let the eigendecomposition be:
\begin{equation}
\mathbf{B} = \mathbf{Q} \, \mathrm{diag}(\boldsymbol{\Lambda}) \, \mathbf{Q}^\top
\label{eq:spectral}
\end{equation}
where $\mathbf{Q}$ contains the eigenvectors as columns and $\boldsymbol{\Lambda} = [\Lambda_1, \dots, \Lambda_d]^\top$ collects the eigenvalues. The matrix logarithm follows by spectral mapping:
\begin{equation}
\ln(\mathbf{B}) = \mathbf{Q} \, \mathrm{diag}(\ln \boldsymbol{\Lambda}) \, \mathbf{Q}^\top
\label{eq:matrix_log}
\end{equation}

Differentiating this operation requires the Fréchet derivative of the matrix logarithm. Using the Daleckii–Kreĭn formula \cite{higham2008functions}, the directional derivative in direction $\mathbf{E}$ is:
\begin{equation}
\mathrm{d} \ln(\mathbf{B})[\mathbf{E}] = \mathbf{Q} \left( \mathbf{L} \odot (\mathbf{Q}^\top \mathbf{E} \, \mathbf{Q}) \right) \mathbf{Q}^\top
\label{eq:frechet}
\end{equation}
where $\odot$ denotes the Hadamard (elementwise) product and $\mathbf{L}$ is the Loewner matrix with entries:
\begin{equation}
L_{ij} = 
\begin{cases}
\dfrac{\ln \Lambda_i - \ln \Lambda_j}{\Lambda_i - \Lambda_j}, & i \neq j \\[8pt]
\dfrac{1}{\Lambda_i}, & i = j
\end{cases}
\label{eq:loewner}
\end{equation}

The off-diagonal terms become ill-conditioned when $\Lambda_i \approx \Lambda_j$. Under large rotations with minimal stretch, the eigenvalues of $\mathbf{B}$ cluster near unity, triggering this singularity. We regularize by switching to the diagonal form $L_{ij} = 1/\Lambda_i$ when $|\Lambda_i - \Lambda_j| < \epsilon$, ($\epsilon = 10^{-12}$ in our experiments), which corresponds to the limiting value as $\Lambda_j \to \Lambda_i$.

\subsection{Algorithm}
\label{sec:method_algorithm}

We summarize the optimization framework, which integrates the differentiable MPM solver with a gradient-based optimizer, below.

\begin{enumerate}
    \item \textbf{Discretization:}
    The computational domain $\Omega$ is discretized into a fixed Eulerian grid and the design domain $\Omega_D$ into a set of Lagrangian material points. The grid provides the computational mesh for the force balance equation solve. The material points serve as quadrature points for stress integration and store the design field.
    
    \item \textbf{Initialization:}
    Each particle is assigned an initial volume $V_p^0$, mass $m_p = \rho V_p^0$, and identity deformation gradient. For single-material design, the pseudo-densities initialize uniformly as $\gamma_p = V^*$. Additionally, for multi-material design, the network weights $\bm{w}$ are initialized via Xavier normal initialization, producing uniform volume fractions $v_s = 1/S$.
    
    \item \textbf{Boundary Conditions:}
    Dirichlet conditions prescribe nodal displacements on $\Gamma_d$.  Point loads are applied through $\mathbf{f}^{\mathrm{ext}}_p$ on material points. 
    
    \item \textbf{Optimization Loop:}
    Each iteration is detailed as follows:
    \begin{enumerate}
        \item \textbf{Design-to-physics mapping:} For pseudo-density, compute $\lambda_p$ and $\mu_p$ via \Cref{eq:simp_single}. For neural network, evaluate $\bm{v}(\bm{x}_p; \bm{w})$ and compute properties via \Cref{eq:simp_interp,eq:density_interp}.
        
        \item \textbf{Forward solve.} At each load step, solve $\mathbf{R} = \mathbf{0}$ via Newton-Raphson until $\|\mathbf{R}\| < \epsilon_{\mathrm{NR}}$. Update particle states: positions $\bm{x}_p$, deformation gradients, stresses $\boldsymbol{\sigma}_p$, and volumes $V_p$.
        
        \item \textbf{Objective and constraint evaluation:} Compute $J_c$ or $J_m$ and $g_v$ or $g_m$ from the converged configuration.
        
        \item \textbf{Sensitivity analysis:} Reverse-mode AD propagates gradients from objectives to design variables. IFT bypasses Newton iteration unrolling. Checkpointing manages memory across load steps.
        
        \item \textbf{Design update:} For pseudo-density, update $\gamma_p$ via Method of Moving Asymptotes (MMA) \cite{svanberg1987method} or optimality criteria. For neural network, update weights $\bm{w}$ via ADAM \cite{kingma2014adam}.
        
        \item \textbf{Continuation scheme:} Increase the SIMP exponent $q$ toward its target value.
        
        \item \textbf{Convergence check:} Terminate if $|(\Delta J_m / J_m) \text{ or } (\Delta J_c / J_c)| < \epsilon_{\mathrm{opt}}$ or the maximum iteration count is reached.
    \end{enumerate}
\end{enumerate}


%% file: 3_result.tex
\section{Numerical Experiments}
\label{sec:expts}

This section presents numerical experiments demonstrating the proposed framework. All simulations are performed on an Apple MacBook M4 Pro using the JAX library \cite{jax2018github}. Unless otherwise specified, the default parameters are as follows:

\begin{itemize}
    \item \textbf{Mesh:} The computational domain is discretized using a structured background grid. The domain sizes ($\Omega$ and $\Omega_D$) and resolution vary by example and are specified in each subsection.
    
    \item \textbf{Material points:} We assign four material points per element per spatial dimension, yielding a total of 16 material points per element in the two-dimensional setting.
    
    \item \textbf{Neural network:} For NN-based optimization problems, the topology is parameterized via a network with a 100-term Fourier projection layer and two hidden layers of 40 neurons each, using a softmax output activation.
    
    \item \textbf{SIMP continuation:} The penalization exponent $q$ is initialized to 1 and incremented by $5 \times 10^{-2}$ per iteration to a maximum of 5.
    
    \item \textbf{Newton-Raphson solver:} For the Newton-Raphson solver, we use a convergence tolerance $\epsilon_{\mathrm{NR}} = 10^{-7}$, a maximum of $50$ iterations, and a backtracking line search (Armijo parameter $c = 10^{-4}$)\cite{armijo1966minimization, nocedal2006numerical}.
    
    \item \textbf{Optimizer:} For single-material problems, we employ the Method of Moving Asymptotes (MMA) \cite{svanberg1987method} with a move limit of $10^{-2}$. For multi-material problems, we use the ADAM \cite{kingma2014adam} optimizer with a learning rate of $10^{-2}$.
    
    \item \textbf{Constraint handling:} For single-material problems, constraints are enforced directly within MMA. For multi-material neural network problems, we formulate the constrained optimization problem using a log-barrier penalty (\cite{Kervadec2019ConstrainedExtensions}) with an initial parameter $\tau_0 = 3$ and growth factor $\mu = 1.02$ is used.
    
    \item \textbf{Convergence:} The optimization terminates when relative objective change $|(\Delta J_m / J_m) \text{ or } (\Delta J_c / J_c)| < 10^{-4}$ or after a maximum number of iterations specified per example.
\end{itemize}

\begin{table}[h]
\centering
\caption{Material properties for single-material examples.}
\label{tab:single_material}
\begin{tabular}{|l|c|}
\hline
\textbf{Parameter} & \textbf{Value} \\ \hline
Lame's lambda $\lambda_0$ & $4.3 \times 10^9$ \\ \hline
Lame's mu $\mu_0$ & $1.1 \times 10^9$ \\ \hline
Mass density $\rho$ & $1600$ kg/m$^3$ \\ \hline
Thickness $t$ & $1.0$ mm \\ \hline
\end{tabular}
\end{table}


\subsection{Validation}
\label{sec:expts_validation}

\begin{figure}[H]
    \begin{center}
\includegraphics[scale=0.5,trim={0 0 0 0},clip]{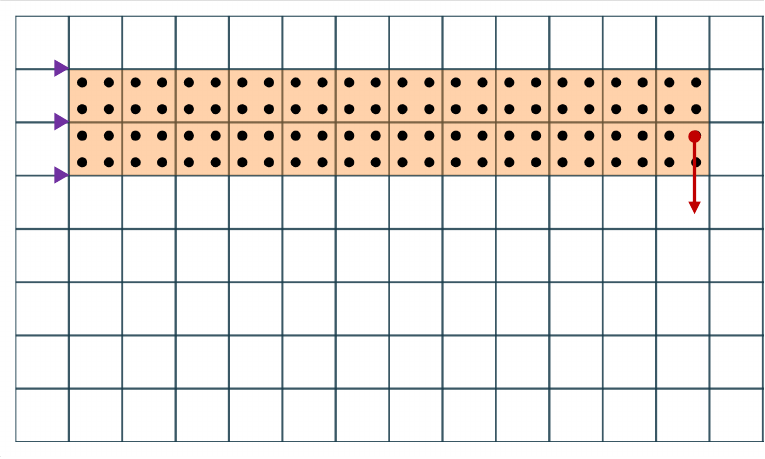}
        \caption{ Representative domain discretization showing the boundary conditions, grid, and material points for a mid-cantilever beam.}
\label{fig:mid_cantilever_domain}
    \end{center}
\end{figure}

We first validate the proposed framework on a TO of a cantilever beam. The setup of the problem is illustrated in \Cref{fig:mid_cantilever_domain}. Here, we consider a grid ($\Omega$) with $80 \times 40$ cells. Furthermore, we consider a rectangular design domain ($\Omega_D$) of dimensions $0.12 \times 0.03$ m that occupies $54 \times 27$ grid cells. Additionally, a Dirichlet condition where the left edge is clamped is imposed on the grid cells while a downward force $f$ is applied at a material point on the mid-right edge of the design domain. The volume fraction constraint is set to $0.5$ (corresponds to an allowed mass of $2.88 \times 10^{-3}$ kg) of the total domain volume $\Omega_D$.

\begin{figure}[H]
    \begin{center}
\includegraphics[scale=0.35,trim={0 0 0 0},clip]{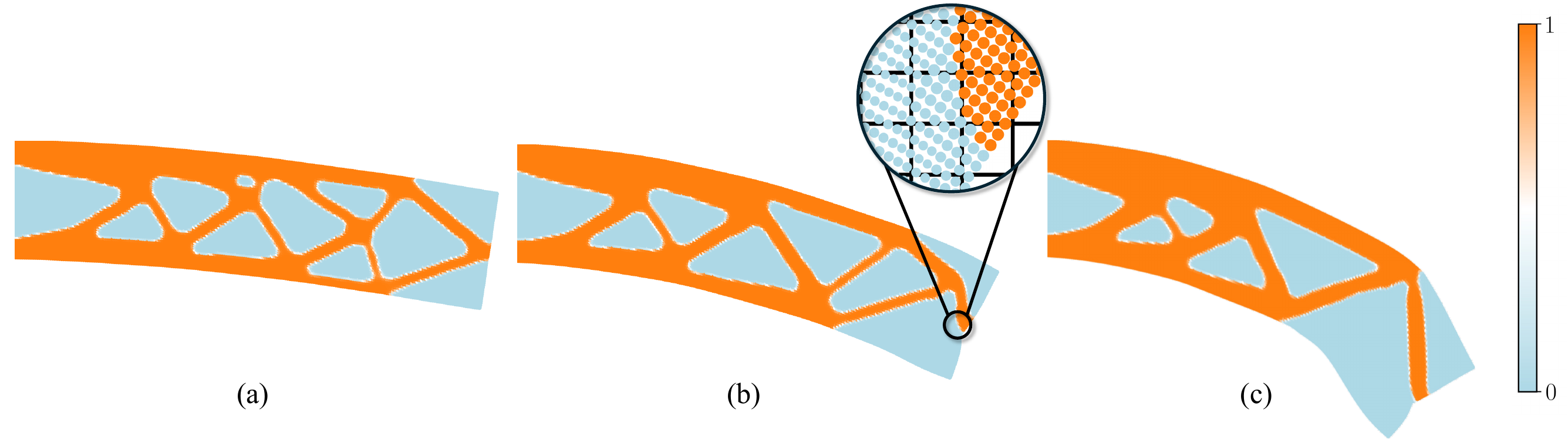}
        \caption{ Optimized designs and corresponding deformed configurations for increasing loads: 
(a) $(f = 0.1\,\mathrm{kN},\; J_c = 1.3)$, 
(b) $(f = 0.25\,\mathrm{kN},\; J_c = 7.6)$, and 
(c) $(f = 0.5\,\mathrm{kN},\; J_c = 24.9)$.}
        \label{fig:val_density}
    \end{center}
\end{figure}

\Cref{fig:val_density} shows the optimized topologies for increasing load magnitudes: $f = 0.1$, $0.25$, and $0.5$ kN. Under the smallest load, the topology is nearly symmetric about the horizontal midplane (\Cref{fig:val_density}(a)). This resembles the solution obtained from linear modeling. As the load increases, the optimized designs become increasingly asymmetric (\Cref{fig:val_density}(b–c)). A single tensile member emerges on the right-hand side of the beam. This member grows in both length and width with increasing load. The emergence of this tensile member reflects the dominant role of geometric nonlinearity at large deflections. After significant deformation, the tensile member provides the most direct load transmission path. As the nonlinear effect intensifies, this member becomes the primary load-bearing component. The final optimized topologies are consistent with solutions reported in existing studies \cite{wang2014interpolation}.

\begin{figure}[H]
    \begin{center}
\includegraphics[scale=0.35,trim={0 0 0 0},clip]{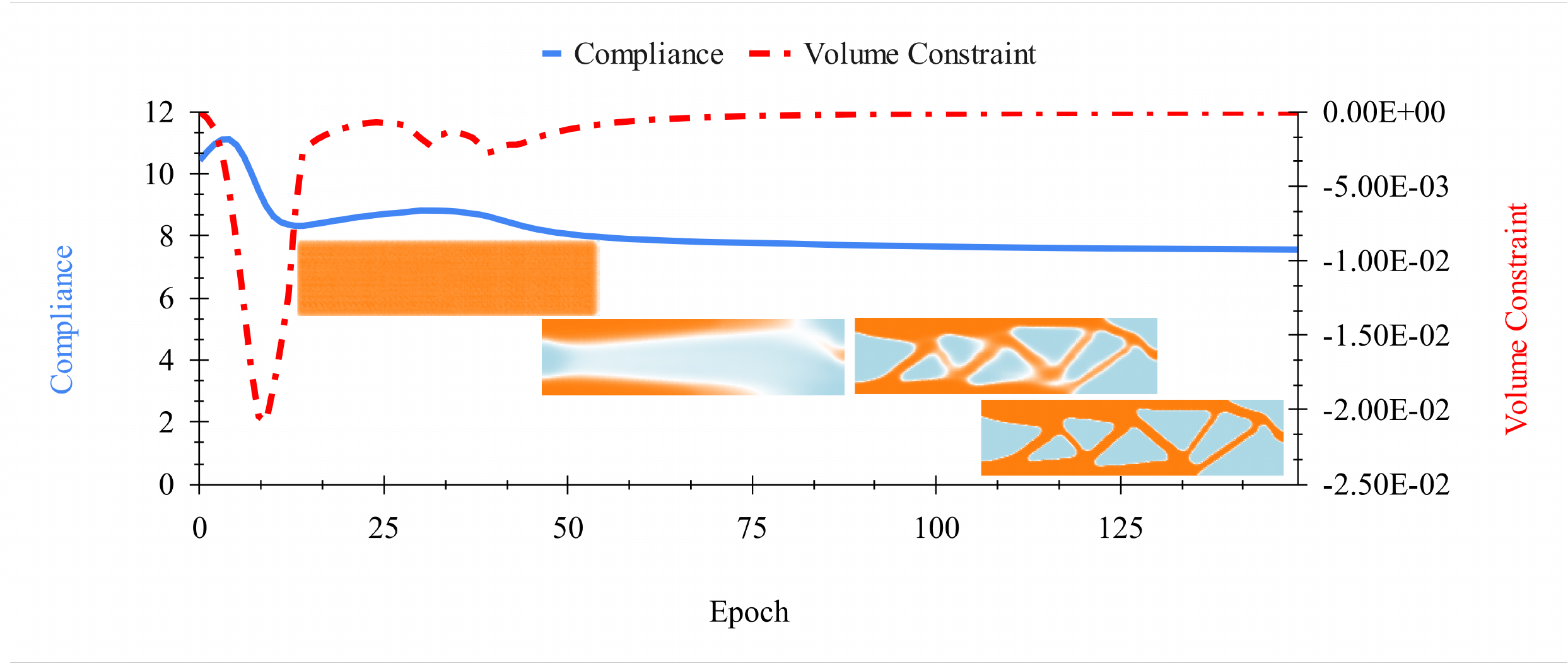}
        \caption{  Convergence of the objective and constraints for a prescribed load of $0.25\,\text{kN}$.}
        \label{fig:convergence}
    \end{center}
\end{figure}

\Cref{fig:convergence} shows the convergence history for the $f = 0.25$ kN case. Both the objective (compliance) and the constraint (volume fraction) exhibit stable convergence. The optimization was completed in 4 minutes and 36 seconds. Similar convergence behavior was observed for the other experiments.

\subsection{Compliance Minimization of Multimaterial Designs}
\label{sec:expts_mmtoComp}

We now extend the cantilever benchmark to multiple materials. The problem setup remains unchanged. Three materials (plus void) with distinct Lame parameters and mass density, as shown in (\Cref{tab:multi_material}) are used. A mass constraint of $M^* = 2.88 \times 10^{-3}$ kg is imposed.

\begin{table}[H]
\centering
\caption{Material properties for multi-material examples.}
\label{tab:multi_material}
\begin{tabular}{|c|c|c|c|c|}
\hline
\textbf{Material} & \textbf{Color} & \textbf{$\lambda$ (GPa)} & \textbf{$\mu$ (GPa)} &\textbf{$\rho$ (kg/m$^3$)} \\ \hline
1 & {\color[HTML]{A86BFF} Violet} & 17.1 & 4.3 & 4000 \\ \hline
2 & {\color[HTML]{FFC700} Yellow} & 4.3 & 1.1 & 1600 \\ \hline
3 & {\color[HTML]{FF4A00} Orange} & 1.4 & 0.35 & 750 \\ \hline
4 & White (Void) & $1.4 \times 10^{-4}$ & $3.5 \times 10^{-5}$ & $10^{-1}$ \\ \hline
\end{tabular}
\end{table}

\begin{figure}[H]
    \begin{center}
\includegraphics[scale=0.35,trim={0 0 0 0},clip]{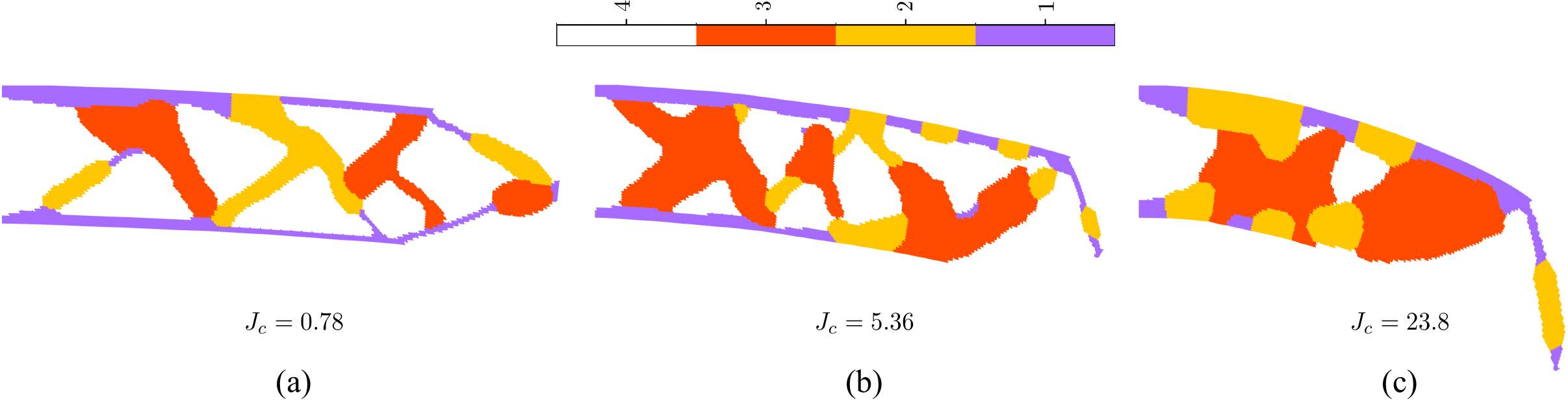}
        \caption{ Optimized design obtained from TO using multiple materials for increasing loads: 
(a) $(f = 0.1\,\mathrm{kN},\; J_c = 0.78)$, 
(b) $(f = 0.25\,\mathrm{kN},\; J_c = 5.36)$, and 
(c) $(f = 0.5\,\mathrm{kN},\; J_c = 23.8)$.}
        \label{fig:multi_mat_comp}
    \end{center}
\end{figure}

\begin{figure}[H]
    \begin{center}
\includegraphics[scale=0.35,trim={0 0 0 0},clip]{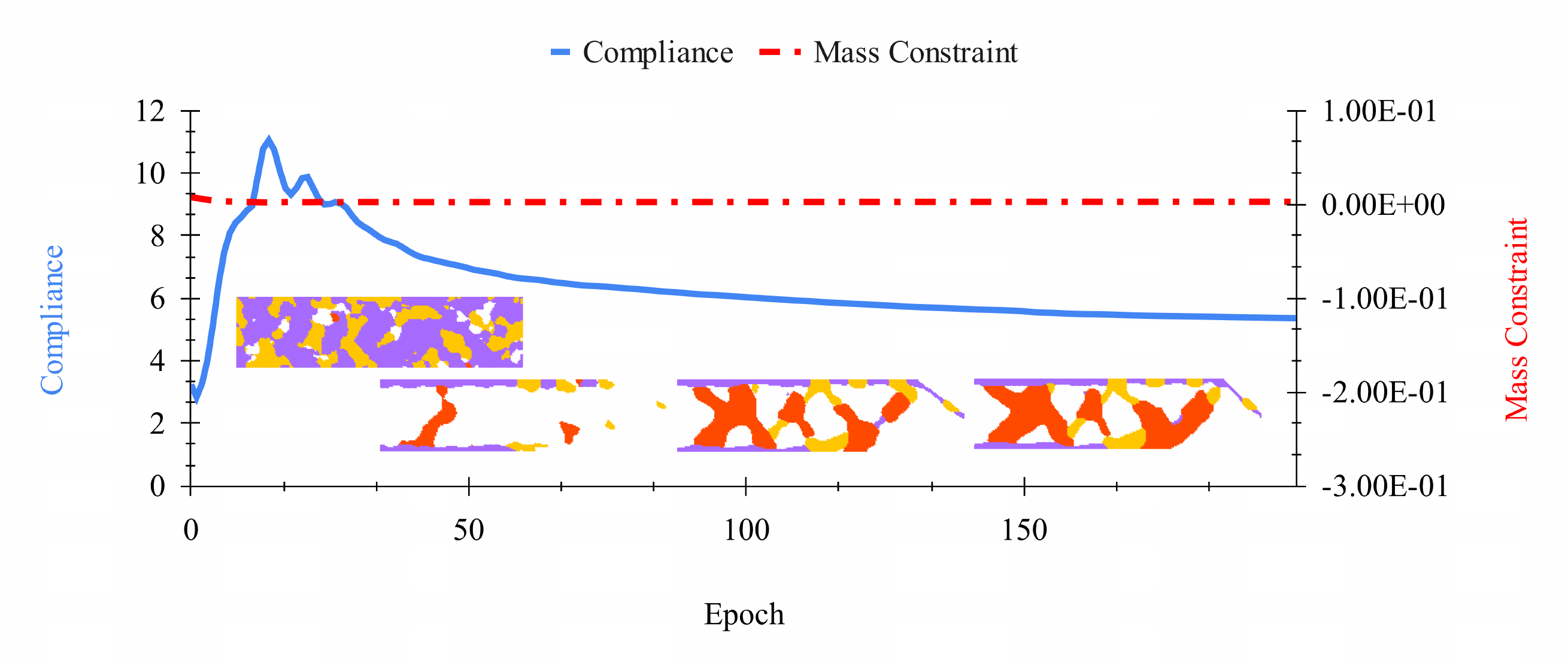}
        \caption{  Convergence of the objective and constraints of multi-material TO for a prescribed load of $0.25\,\text{kN}$.}
        \label{fig:convergence_mmto}
    \end{center}
\end{figure}

\Cref{fig:multi_mat_comp} shows the optimized multi-material designs for each load case. Compared with the corresponding single-material designs, the multi-material formulation achieves lower compliance for the same allowable mass due to the additional design freedom provided by multiple materials. Furthermore, \Cref{fig:convergence_mmto} shows the convergence history for the $f = 0.25$ kN case. Both the objective (compliance) and constraint (mass constraint) exhibit stable convergence. The optimization completed in 5 minutes and 12 seconds. Similar convergence behavior was observed for the other experiments.

\subsection{Design of Grippers}
\label{sec:expts_grippers}

The design of soft robotic grippers requires both flexibility for grasping and stiffness for load-bearing. We demonstrate the proposed framework on the design of compliant grippers \cite{du2025topology}, using the material properties listed in \Cref{tab:multi_material}.

\begin{figure}[h!]
    \begin{center}       \includegraphics[scale=0.45,trim={0 0 0 0},clip]{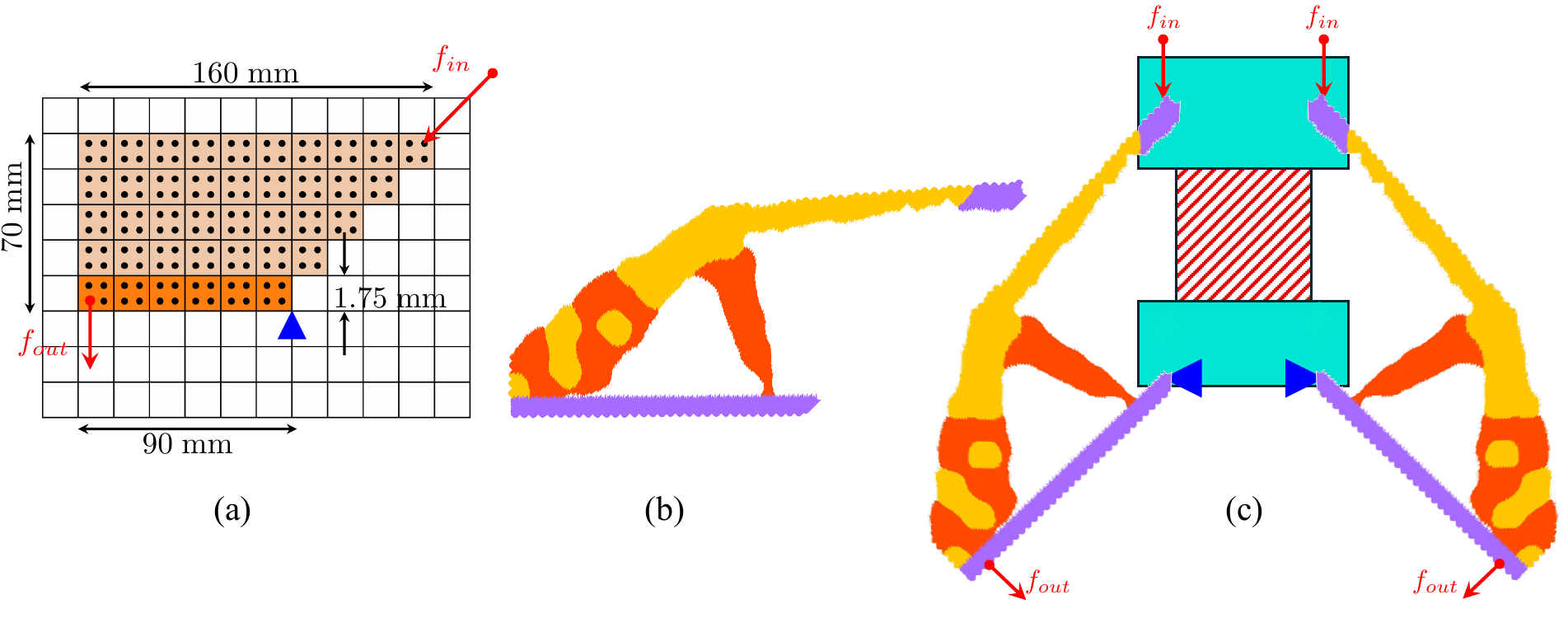}
        \caption{ Design of compliant grippers. (a) Representative design domain, non-design contact region, boundary conditions, and input/output ports for a single gripper. (b) Optimized gripper topology obtained using the proposed framework. (c) Assembled gripper configuration under the applied input forces $f_\mathrm{in}$.}
        \label{fig:gripper}
    \end{center}
\end{figure}

\Cref{fig:gripper}(a) shows the design domain and boundary conditions. The domain is a trapezoid with a horizontal extent of 160 mm and a height of 70 mm. A solid non-design region of dimensions $90 \times 1.75$ mm spans the bottom edge; this region represents the contact surface that interacts with grasped objects. The top-right corner receives an input force $f_\mathrm{in}$, while a fixed support constrains the bottom edge near the output port. The objective is to maximize the output displacement $f_\mathrm{out}$ at the left corner of the contact surface for a prescribed input.

\Cref{fig:gripper}(b) shows the optimized topology and \Cref{fig:gripper}(c) shows the assembled configuration under the applied load.
 
%% file: 4_conclusion.tex
\section{Conclusion}
\label{sec:conclusion}

In this work, we present MOTO, a material point method (MPM)-based topology optimization (TO) framework for the design of structures undergoing quasi-static large deformations. In particular, we utilize the implicit Generalized Interpolation Material Point (GIMP) variant of MPM to resolve mesh distortion issues inherent to conventional Lagrangian finite element methods. We incorporated a Hencky hyperelastic constitutive model to capture the response of structures in the finite-strain regime. Furthermore, we leveraged an end-to-end differentiable pipeline implemented in JAX to automate the computation of sensitivities. We presented an optimization framework that supports both single-material pseudodensity and multi-material coordinate neural network design representations. The effectiveness of the framework was demonstrated through validation studies and the design of compliant mechanisms, specifically a soft robotic gripper.

The present study offers several avenues for future research. First, the current formulation focused exclusively on hyperelastic materials. However, many engineering applications involve structures that undergo large plastic deformation, such as energy-absorbing devices. The extension of the framework to include elastoplasticity is of significant interest. Second, while we considered compliance and the design of compliant mechanisms, incorporating stress constraints within the large-deformation MPM context will be essential to prevent material failure. Third, future work should focus on extending soft robotic systems to involve coupled design and actuation and optimizing for dynamic trajectories. Fourth, extending the framework to 3D configurations is necessary for practical applicability. This will require leveraging efficient GPU-accelerated implementations such as \cite{zhao2026geowarp} to address computational demands. Fifth, given that MPM naturally handles contact, extending the framework to design contact structures such as \cite{frederiksen2025TOContact} presents an interesting avenue. Sixth, investigating the imposition of manufacturing constraints within the context of material point-based design representations is a necessary future step. Seventh, MPM's hybrid Lagrangian-Eulerian formulation is well suited to problems involving large topological changes, including crack initiation and propagation, dynamic fracture \cite{nairn2003material, kakouris2017phase}, and impact-driven fragmentation \cite{de2020material}, making it a promising direction for future TO research. Finally, it is of interest to validate optimized designs experimentally by fabricating and testing soft robotic prototypes.